\documentclass[aps,article,groupedaddress]{revtex4}
\usepackage{graphicx}
\begin{document}
\bibliographystyle{apsrev}

\newcommand{\be}{\begin{eqnarray}}
\newcommand{\ee}{\end{eqnarray}}
\newcommand{\ds}[1]{
#1{\hskip-2.0mm}/
}
\newcommand{\lb}[1]{
\label{#1}
}

\title{Systematic Study of the Single Instanton Approximation in QCD}

\author{P. Faccioli, E. V. Shuryak}
\email{faccioli@tonic.physics.sunysb.edu,shuryak@dau.physics.sunysb.edu}
\affiliation{Department of Physics and Astronomy\
State University of New York at Stony Brook/
Stony Brook, New York 11794, USA}
\date{\today}


\begin{abstract}

Single-instanton 
approximation (SIA) is often used to evaluate analytically
instanton contributions
 euclidean correlation function in QCD at small distances.
We discuss how this approximation can be
  consistently derived from the theory of  
instanton ensemble and give precise definitions to a number of 
different ``quark effective masses'',  generalizing the parameter 
$m^*$, which was  introduced long ago to account for the 
collective contribution of the whole ensemble. 
We test numerically the range of applicability of the SIA
for different quantities. Furthermore, 
we determine all the effective masses (for random and interacting
instanton liquid models) as well as from phenomenology,
and discuss to what extent those are universal.lan
\end{abstract}
\pacs{xxx}
\maketitle

\section{Introduction}
\label{intro}

The instanton liquid model  of the QCD vacuum \cite{shuryak82} 
is based on a
semiclassical approximation, in which 
 all gauge configurations are replaced by 
an ensemble of topologically non-trivial fields, instantons 
and anti-instantons. It remains a model because
we do not still understand why large-size instantons are not present
in
the ensemble.
Fits to phenomenology and later lattice studies  showed that
 their total density is $n_0\simeq 1 fm^{-4}$ while the typical
size of about $\rho\sim 1/3 fm$,
leading to small diluteness parameter  $n_0\rho^3\sim 10^{-2}$
 \cite{shuryak82}.
With these parameters, the model quantitatively explains such
important phenomena as spontaneous $SU(N_f)$ 
chiral symmetry breaking for $N_f$ quark flavors, 
the explicit U(1) symmetry breaking, and 
many more other details of hadronic correlators and spectroscopy
 (for a recent example see discussion of vector and axial correlators
\cite{SS_01}, for a
 review see \cite{shuryakrev}). 
  The main feature of the instanton\footnote{For simplicity, we shall often use the term ``instanton'' to denote 
instantons and/or 
anti-instantons.}
 ensemble is that each  pseudo-particle
is an effective vertex with $2N_f$ quark lines \cite{thooftzm1},
which are exchanged between them and fill the vacuum.
A theory is developed, called Interacting Instanton Liquid Model
(IILM) which include these 't Hooft interactions {\em to all orders}
 \cite{shuryakrev}.

If new sources (external currents) are added, they produce 
extra quarks which
interact with those in vacuum and
produce non-trivial correlation functions. In particular,
many (Lorentz scalar) chirally odd local  operators obtain non-zero 
vacuum expectation values.
In general, all of those 
 ``condensates'' and   correlation functions are determined by
the interaction of instantons and thus 
 depend on the global (collective)  
properties of the ensemble.

 On the other hand, as the instanton vacuum is 
fairly dilute, one may think that the correlation  
functions at distances short compared to instanton spacing $x\ll R=n^{-1/4}\sim 1 fm$
may be dominated by a $single$ instanton, the  closest (or leading) one 
(LI).
This framework ( which we shall refer to as the 
``single instanton approximation'', SIA ) has the advantage to allow 
to carry out calculations analytically.
It is therefore possible to obtain closed expressions 
for instanton  contribution to Green's functions
 in momentum or in Borel space.

In SIA  
 collective contribution of all
 instantons {\em other than the leading one}
are taken care of by a single effective
parameter, usually called effective mass, $m^*$. In the simplest
approximation, it can be associated with an $average$ value of the
quark condesate  \cite{shifman80}:
\be
\label{mstar}
m^*= m - \frac{2}{3}\pi^2\,\rho^2 \,<\bar{u}\,u>,
\ee
 which leads to 
the value  $m^*\simeq 170 MeV$ \cite{shuryak82}.
Note that it is already very different from what one infers from the same 
model for long distance 
(or zero Euclidean momentum) limit of the quark propagator, which
gives {\em constituent quark mass} of the order of 400 MeV.

Furthermore,
although the SIA has been used in several phenomenological
 studies (e.g. \cite{shuryak82}, \cite{forkel,forkel2,forkel3},
 and references therein), its derivation 
 was never discussed in detail,  its range of applicability
was never quantitatively checked, and the values of relevant effective
masses
well specified.
And
indeed, if one uses  the value  $m^*\simeq 170 MeV$ the
 correlation functions, evaluated in the SIA,
do not agree with the results of the random and interacting 
instanton liquid \cite{shuryakrev}.

In this paper we identify the origin of such discrepancy
and calculate the values of effective mass appropriate for different
observables. This analysis reveals that the discrepancy between
SIA and full liquid calculations is due to an incorrect estimate
of the effective mass, $m^*$.
We also
present a systematic study of the SIA in QCD by itself.
We show that the approach is really accurate 
only for calculations that involve operators of dimension six or more,
or correlators with
more than one zero-mode propagator.
We shall also prove that the mass terms, appearing 
in matrix elements involving different
numbers of zero-mode propagators, are indeed independent parameters that have
to be fixed separately.
We provide with the definitions of all such mass factors in 
terms of averages of the instanton ensembles
 and prove that they are nearly universal,
 i.e. the same for all similar correlation functions.

  
The paper is organized as follows. 
In section \ref{theory} we derive the SIA from the theory of the
instanton ensemble,
 in section \ref{reliability} we 
 present the results of our numerical simulations that 
estimate the contribution from the leading-instanton to several 
correlation functions.
In section \ref{mass}, we evaluate the effective mass terms both from the
 random and interacting instanton liquid and compare it
 with the values obtained phenomenologically from the 
pion sum-rule.
In section \ref{determass} we compare our effective masses with the 
so-called ``determinantal masses'', which are other effective parameters that
can be defined in  terms of averages of
 the fermionic determinant.
The main results of our analysis are summarized in 
section \ref{conclusions}.
   

\section{Quark Propagator}
\label{theory}

In this section we review how the quark propagator in the instanton vacuum
 is obtained and present  consistent derivation of the SIA.

The quark propagator in general background field is
\begin{eqnarray}
\label{propdefinition}
S_I(x,y)=<x|(i\ds{D}_I+i\,m)^{-1}|y>,
\end{eqnarray}
where $\ds{D}_I$ denotes the Dirac operator. The inverse
(\ref{propdefinition}) can be formally represented as an 
expansion in eigenmodes of the Dirac operator:
\be
\label{propspectral}
S_I(x,y) = \sum_\lambda \, \frac{\psi_\lambda (x)\, \psi^\dagger_\lambda(y)}
	{\lambda+ i\, m}, \qquad i\,\ds{D}_I\psi_\lambda(x) 
	= \lambda \psi_\lambda(x).
\ee
From eq. (\ref{propspectral}) it follows that the propagator of light quarks
is dominated by eigenmodes with small virtuality.

We begin by considering the academic case in which the
vacuum contains only one isolated instanton.
One eigenmode of $\ds{D}_I$
with zero virtuality  (zero-modes)  is given by 't Hooft
 \cite{thooftzm1}, \cite{thooftzm2}:
\be
\label{zmodes}
i\ds{D}\,\psi_0(x) &=& 0, \nonumber\\
\psi_{0\,\,a\,\nu}(x;z) = \frac{\rho}{\pi}
\frac{1}{((x-z)^2+\rho^2)^{3/2}}&\cdot&\left[
\frac{1 - \gamma_5}{2}\, \frac{\ds{x}-\ds{z}}{\sqrt{(x-z)^2}}
\right]_{\alpha\,\beta} U_{a\,b}\,\epsilon_{\beta\,b},
\ee
where $z$ denotes the instanton position, 
$\alpha, \beta= 1,\cdots 
4$ are spinor indices and $U_{a b}$ represents a 
general group element.

Isolating the contribution from zero-modes we can write:
\be
\label{S_I}
S_I(x,y;z) & = &\frac{\psi_0(x-z)\,\psi^\dagger_0(y-z)}{i\,m} + 
	\sum_{\lambda\ne 0} \, \frac{\psi_\lambda (x-z)\,
	 \psi^\dagger_\lambda(y-z)}{\lambda + i\, m}\nonumber\\
	& = & S^{zm}_I(x,y;z) + S^{nzm}_I(x,y;z).
\ee
The zero-mode part of the propagator in the field of one 
instanton can be evaluated from (\ref{S_I}) and
(\ref{zmodes}) to give, \cite{dyakonov86}:
\be
\label{S_zmI}
S^{zm}_I(x,y;z)= 
\frac{(\ds{x}-\ds{z})\gamma_\mu \gamma_\nu 
(\ds{y}-\ds{z})}{8 m}
\left[
\tau_\mu^- \tau_\nu^+ \,\frac{1-\gamma_5}{2}
\right]
\, \phi(x-z)\, \phi(y-z),
\ee
where
\be
\phi(t):=\frac{\rho}{\pi}\frac{1}{|t|\,(t^2+\rho^2)^{3/2} },
\qquad \tau_\mu^{\pm} := ({\bf \tau}, \mp i   )
\ee
The corresponding expression in the field of one anti-instanton is obtained 
through  the substitution:
\be
\frac{1-\gamma_5}{2}\longleftrightarrow   \frac{1+\gamma_5}{2}   
\qquad 
\tau^- \longleftrightarrow \tau^+.
\ee

In the chiral limit, $m\to 0$, the expression for $S^{nzm}_I(x,y;z)$
 is also  known exactly \cite{brown78}.
In the limit of small distances ($|x-y|\to 0$), or if the instanton 
is very far away ($|x-z|\to \infty$) one has:
\be
\label{nonzm}
S^{nzm}_I(x,y;z) \simeq S_0(x,y),
\ee
where $S_0$ denotes the free propagator. 
Typically, corrections to eq. (\ref{nonzm}) lead to small contributions
and will be neglected in what follows.
Once the propagator has been calculated, one can in principle evaluate any 
correlation function in the single-instanton background.

Now, let's turn to the realistic vacuum of QCD.
Here, any configuration with a non-zero net topological charge 
would be highly disfavored by the small value of the $\theta$-angle.
Therefore, one is lead to picture the vacuum as an ensemble with equal 
density of instanton 
and anti-instantons.
If the vacuum is dilute enough, the classical back-ground field can 
be approximatively taken to be a superposition of 
separated instantons and anti-instantons \footnote{For sake of simplicity, we 
are explicitly dropping all collective coordinates, except the 
instanton position $z_i$; moreover the use of the singular gauge 
is assumed everywhere.}:
\begin{eqnarray}
\label{sumansatz}
A_\mu(x,\{\Omega_i\}_i)=\sum_{I}A^I_\mu(x,\{\Omega_i^I\}_i)
+\sum_{A}A^A_\mu(x,\{\Omega_i^A\}_i),
\end{eqnarray}
where $\{\Omega_i\}_i$ denotes the set of all  collective coordinates.

The propagator in such back-ground field can then be evaluated as follows
\cite{dyakonov86}. Let's consider the expansion:
\begin{eqnarray}
\label{Sexp1}
S= S_0 + S_0\,\ds{A}\,S_0 + S_0\,\ds{A}\,S_0\,\ds{A}\,S_0 + ...,
\end{eqnarray}
where  integrations over  
the positions of each background field insertion is understood.
The  series (\ref{Sexp1}) can be rearranged so that all terms depending on
the collective coordinates of one
instanton field only are summed up first, followed by all terms depending 
on two instantons and so on. One gets:
\be
\label{Sexp2}
S= S_0 + \sum_I ( S_I - S_0 ) + \sum_{I\ne J} ( S_I - S_0) S_0^{-1} (S_J-S_0)
+...,
\ee
where  $S_I$ denotes the full propagator in the field of 
the instanton $I$ so, in
the approximation ($\ref{nonzm}$) one has:
\be
\label{S_I-S_0}
(S_I-S_0)_{i\,j}(x,y)	\simeq \frac{\psi^I_{0\,i}(x)\, 
\psi^{\dagger\,I}_{0\,j}(y)}{i\,m},
\ee
where we have dropped all collective coordinates indices.
Inserting (\ref{S_I-S_0}) in (\ref{Sexp2}) and dropping also all
spinor indices we get:
\be
S(x,y) & \simeq &  S^0(x,y) +
\sum_I \frac{\psi_0(x)\,\psi^\dagger_0(y)}{i\,m}
 +  \nonumber\\
& &\sum_{I,\,J} \frac{\psi_{0\,I}(x)}{i\,m}
\left(
\int d^4 z \psi^\dagger_{0\, I}(z)(i\ds{\partial}_z +i\,m) 
\psi_{J\,0}(z) - i\,m \delta_{I\,J}
\right)
\frac{\psi_{0\,J}^\dagger(y)}{i\,m}+...,
\ee
where $- i m\, \delta_{I,J}$ has been added in order to relax the 
$J\ne I$ constraint in the summation.
All the terms, starting from the second on, form a geometrical progression, 
which can be re-summed to give:
\be
\label{S}
S(x,y) & \simeq &  S^0(x,y) 
 + \sum_{I,\,J}
\psi_{0\,I}(x)\,
\left(
\frac{1}{T
+ o(m)}
\right)_{I J}
\psi_{0\,J}^\dagger(y),
\ee
where $T_{I J}$ denotes the overlap matrix in zero-modes subspace
\be
T_{I J}=\int d^4 z \psi^\dagger(z)_I(i\ds{\partial})\psi(z)_J .
\ee

In (\ref{S}), the zero-mode part the quark propagator is approximatively 
written as a
bilinear form in the space spanned by the quark zero-mode
wave functions.
From (\ref{zmodes}) it follows that the contribution coming
from all the terms in the sum 
associated to instantons very far away from the points $x$ and $y$ will be 
negligible.
In particular, the biggest term in (\ref{S}) is associated to
the closest instanton, $I^*$.
Such instanton is dominating if the average of the correlation
function calculated retaining only the $(I^*,I^*)$ term in (\ref{S})
is much larger than
the average of the same quantity calculated from all other terms in 
the sum (\ref{S}). Notice that this is a much weaker assumption than 
demanding
\be
\psi_{0\,I^*}(x)\,
\left(
\frac{1}{T + o(m)}
\right)_{I^* I^*}
\psi_{0\,I^*}^\dagger(y)
\gg \sum_{I\ne I^*,\,J\ne I^*}
\psi_{0\,I}(x)\,
\left(
\frac{1}{T + o(m)}
\right)_{I J}
\psi_{0\,J}^\dagger(y),
\ee
for \emph{each} configuration.

Let us summarize the framework developed so far.
First of all,  the inverse matrix 
$\left( 
\frac{1}{T}
\right)_{I J}$
contains all the information about the particular configuration of the 
instanton ensemble. In order to evaluate  correlation functions, 
one needs to average over all possible configurations.
Since contributions from distant instantons are suppressed by their 
zero-mode wave functions, one expects correlation 
functions with the
highest number of zero-modes to be most influenced by the
leading-instanton $I^*$.
If it is possible to retain only 
the contribution from $I^*$, the global properties of the ensemble 
are present in the
 matrix element 
$\left( 
\frac{1}{T}
\right)_{I^* I^*}$.

As it was suggested long time ago by one of us \cite{shuryak82}, 
one can represent collective contribution of all other instantons,
by introducing an effective mass associated to quark propagating in the 
zero-modes.
In other words, one  assumes that for $|x-y|< 1 fm$, 
the quark propagator can be written as:
\be
\label{SIAprop}
S(x,y)=\frac{\psi_0(x)\psi_0^\dagger(y)}{i m^*}.
\ee
With such propagator
all  quark correlation functions
in the instanton  back-ground could  be evaluated simply by computing
 all relevant Feynman diagrams and then 
averaging over the instanton  collective coordinates
\footnote{Notice that for all gauge invariant matrix elements, 
the average over the color orientation is trivial.}.

More specifically, in Random Instanton Liquid Model (RILM) 
 one introduces a model instanton density $n(\rho)$, 
\be
n_I(\rho):= \bar{n}_I \,d(\rho), 
\ee
where  $\bar{n}_I=\bar{n}_A\simeq \frac{1}{2} fm^{-4}$ and
$d(\rho)$ represents the instanton size distribution. The latter is 
 schematically taken  to be:
\be
\label{deltarho}
d(\rho)=\delta(\rho-\bar{\rho}).
\ee
with $\bar{\rho}\simeq 1/3 fm$. 
This approach has the advantage to be considerably simple and was also
proven to be quite phenomenologically successful \cite{forkel,forkel2}.
However,  we show below that 
the effective mass defined in
(\ref{SIAprop})  is a quantity quite different from its naive estimate
(\ref{mstar}).

In order to clarify the statement, let us first consider the quark condensate:
\be
\label{uu}
\chi_{uu}= <0|Tr \,\bar{u}(x) u(x) |0> = <Tr \,S(x,x)>,
\ee  
where, in general, the average is done over all possible gauge
field configurations. 
In the SIA is easily evaluated: 
\be
<0| \bar{u} (x) \,u(x) |0>= 
\int d^4 z \,\int d\rho\, \bar{n} \,d(\rho)
\left[
\frac{-2\,\rho^2}{[(z-x)^2+\rho^2]^3\,\pi^2\,m_{uu}},
\right]
\ee
where, for reasons that will become clear shortly, we have denoted with
 $m_{uu}$ the quark effective mass and 
$\bar{n}:= \bar{n}_I +\bar{n}_A$.
After performing the integrations one finds:
\be
\label{uuSIA}
\chi_{uu}= -\frac{\bar{n}}{m_{uu}},
\ee
for any normalized $d(\rho)$.  

Now, repeating the same calculation in the full liquid
\footnote{Here, we have neglected all 
small current quark mass
terms in $1/T$.} gives:
\be
\label{uuRIL}
\chi_{uu} = 
\left< 
Tr 
\left[
\sum_{I,J} \psi_{0\,I}(x) \, 
\left(
\frac{1}{T}
\right)_{I\,J}
\psi_{0\, J}^\dagger(x)
\right]
\right>,
\ee
where, again, the average is made over all possible configurations of the 
ensemble.
A comparison between (\ref{uuSIA}) and (\ref{uuRIL}) gives:
\be
\label{mstar1}
m_{uu}:=-\frac{\bar{n}}
{ 
\left< 
Tr 
\left[
\sum_{I,J} \psi_{0\,I}(x) \, 
\left(
\frac{1}{T}
\right)_{I\,J}
\psi_{0\, J}^\dagger(x)
\right]
\right>
}
\ee

Let us now consider another quark condensate:
\be
\label{uudd}
\chi_{uudd}:= <0| Tr\,[ \bar{u}(x) u(x) ]\cdot Tr\,[ \bar{d}(x) d(x) ]  |0>
= < [ \,Tr\, S(x,x)]^2>.
\ee
Such condensate receives double contribution from zero-modes. In the SIA one
obtains:
\be
\chi_{uudd}=
\int d\rho \, d(\rho) \frac{\bar{n}}{5\,\pi^2\,\rho^4\,m_{uudd}^2},
\ee
where we have now denoted with $m_{uudd}$ the quark effective mass.

Comparing, as before, with the result of full liquid calculations leads to:
\be
\label{mstar2}
m^2_{uudd}=
\left(\int\,d\rho\, d(\rho)
\frac{\bar{n}}{5\, \pi^2\,\rho^4} 
\right)
\frac{1}{ 
\left<
\left[
Tr
\sum_{I,J} \psi_{0\,I}(x) 
\left(
\frac{1}{T}
\right)_{I\,J}
\psi_{0\, J}^\dagger(x)
\right]^2
\right>
}
\ee

Now, if the effective mass is universal, 
$(m_{uu})^2 = m^2_{uudd}$, it would imply:
\be
\label{universality}
\frac{
\left< 
Tr 
\left[
\sum_{I,J} \psi_{0\,I}(x) \, 
\left(
\frac{1}{T}
\right)_{I\,J}
\psi_{0\, J}^\dagger(x)
\right]
\right>^2}
{\left<
\left[
Tr
\sum_{I,J} \psi_{0\,I}(x) 
\left(
\frac{1}{T}
\right)_{I\,J}
\psi_{0\, J}^\dagger(x)
\right]^2
\right>}=
\frac{5\,\pi^2 \bar{n}}{\int\,d\rho\, d(\rho)\frac{1}{\rho^4}}\simeq
5\,\pi^2 \bar{n}\bar{\rho}^4\sim \frac{5}{8},
\ee 
where we have used the ansatz (\ref{deltarho})
\footnote{Alternatively, we repeated the calculation 
using a parameterization of the lattice 
measurements of $d(\rho)$, which is peaked about somewhat higher
values of $\rho$ ($\rho\simeq 3.9$). 
Both calculation give basically the same result.}.
Some comments on eq. (\ref{universality}) 
are in order.
First of all, in general quark condensate is rather inhomogeneous,
and for parametrically dilute instanton ensemble
this ratio is small. However, with empirical diluteness it happens to
be not so small, about 0.6 . In principle,
by measuring the left-hand side and right-hand-side
of  (\ref{universality})  on the lattice $separately$,
 one can  estimate  the accuracy of the 
universality of the effective mass.

However, since different configurations and even points have different
leading instanton, the corresponding value $T_{I^* I^*}$ fluctuates,
and the average of its different powers in general leads to different
effective masses. (This effect should not be confused with
the inhomogeneity of the condensates discussed above.)
Let us define a parameter $R_m$, such that $R_m=1$ means universal mass
 $(m_{uu})^2 = m_{uudd}^2$: 
\be
\label{Rm}
R_m:=
\frac{
\left< 
Tr 
\left[
\sum_{I,J} \psi_{0\,I}(x) \, 
\left(
\frac{1}{T}
\right)_{I\,J}
\psi_{0\, J}^\dagger(x)
\right]
\right>^2}
{5\,\pi^2\,\bar{\rho}^4\, \bar{n}
\left<
\left[
Tr
\sum_{I,J} \psi_{0\,I}(x) 
\left(
\frac{1}{T}
\right)_{I\,J}
\psi_{0\, J}^\dagger(x)
\right]^2
\right>}.
\ee 


\section{Numerical study of the Single Instanton Approximation.}
\label{reliability}

In general, reliability of the  SIA 
depends on the vacuum diluteness.
In this section, we want to establish whether the
QCD vacuum with realistic density is actually
dilute enough for the leading-instanton to be dominant, at least for 
some observables.

For this purpose we have performed numerical analysis of several correlation 
functions, measured in the random instanton liquid model.
In such ensemble, the vacuum expectation values are obtained by averaging 
over configurations of 
randomly distributed instantons of size $\rho = 1/3 fm$. 
The contribution from the leading-instanton is evaluated by retaining 
only the largest term in (\ref{S}), for each configuration.

We begin by considering two quark condensates $\chi_{uu}$ and
$\chi_{uudd}$, introduced in  (\ref{uu}) and (\ref{uudd}).
We will show later that they represent all generic
observables which   receive contribution from
 one and two   zero-mode propagators, respectively.

In this calculation we average
 $5000$ configurations of 20 instantons in
a box of volume $3.4\times 1.8^3  fm^4$.
The results of this simulation
are presented in table \ref{condensates}.

  \begin{table}
  \caption{Quark condensates evaluated in the full instanton ensemble
and from the leading-instanton, only.}
 \label{condensates}
\begin{tabular}{|c|c|c|}\hline
\multicolumn{1}{|c|}{condensate} & 
\multicolumn{1}{|c|}{complete calculation} &
\multicolumn{1}{|c|}{LI } \\ \hline \hline
$\chi_{uu}$ & $ ( - 232 \pm 5 MeV )^3 $& $(- 198 \pm 1 MeV)^3$ \\ \cline{2-3}
$\chi_{uudd}$ &$ ( 310 \pm 7 MeV)^6 $& $ (309  \pm 3 MeV)^6$ \\ \hline 
 \end{tabular}
 \end{table}

From these results one can see how the accuracy of SIA
(keeping only the closest instanton) depends on the particular matrix 
element being evaluated.
Naturally, the accuracy increases with the dimension of the operator
involved, because it diminish the contribution of distant instantons. 
Specifically, SIA for dimension-six local operators which
receive  contribution from two zero-mode propagators
 agree with full calculation 
within a few percent.
On the other hand, prediction for operators/correlators with 
only one zero-mode propagators are not really accurate: the 
error in quark condesate is large ( $\gtrsim 35\%$ ).

Next we  consider two-point correlation functions.
This allows us to determine the scale at which the closest instanton is no 
longer dominant.
At this purpose we have measured the pion pseudo-scalar two point function,
\be
\label{pion2point}
P(x) := <0| J_5(x) J^\dagger_5 (0)|0>,
\ee
where,
\be
J_5(x) := \bar{u}(x)\,\gamma_5 \,d(x).
\ee

\begin{figure}[h|t]
 \includegraphics[height=7.0cm,width=8.0cm]{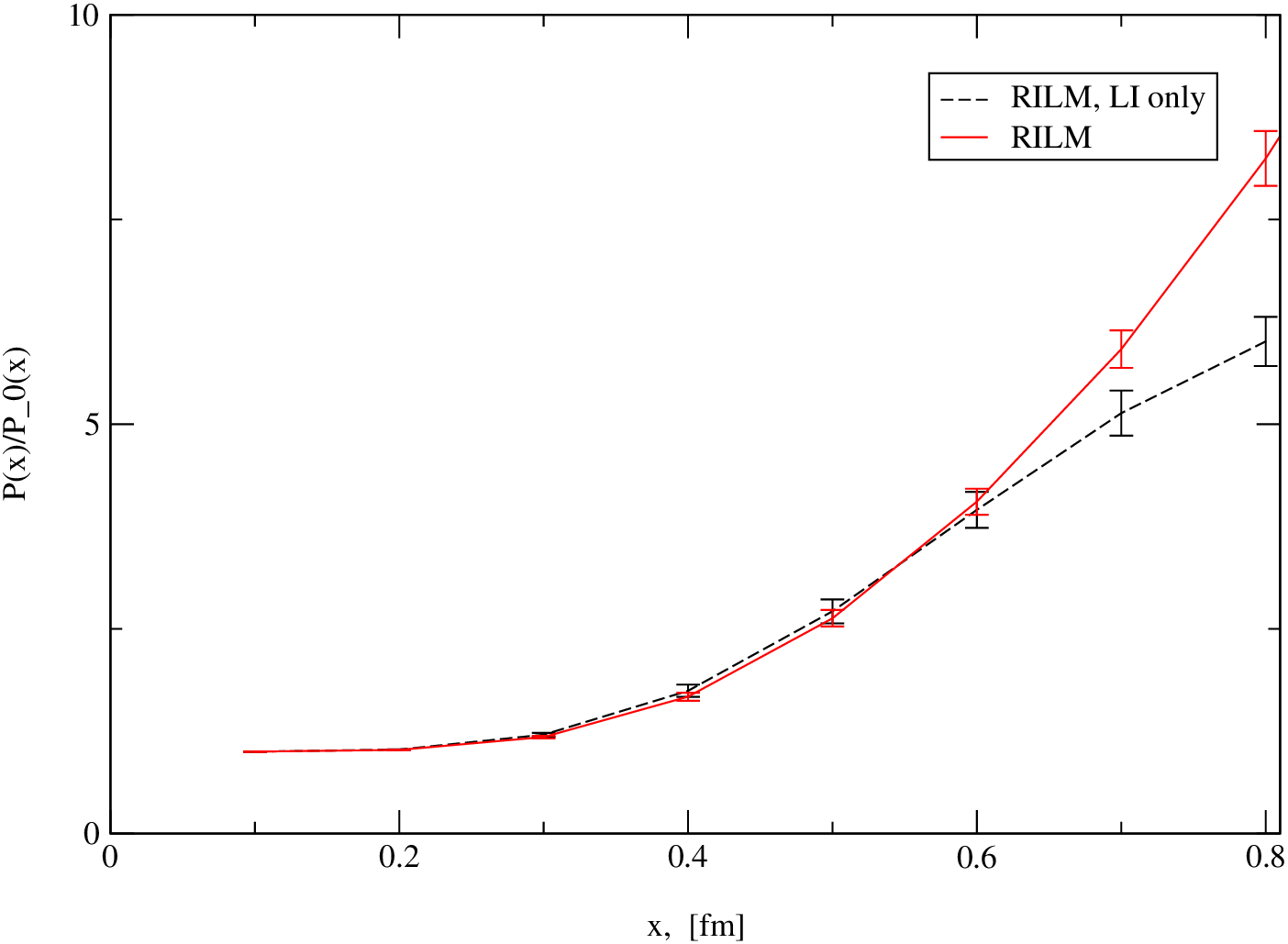}
 \caption{Pion pseudo-scalar correlation function in the RILM,
normalized to the same correlation function in the free theory.
The solid line corresponds to the full RILM simulation, 
the dashed line denotes the leading-instanton contribution.}
 \label{RILvsLI}
 \end{figure}

This particular choice is motivated by the fact that such correlation 
function is known to receive maximal contribution from quark zero-modes
 \cite{shuryakcorr}.
One expects many instantons effects to become important for $|x|$ 
larger than the instanton size 
and smaller than the typical distance between two neighbor instantons:
\be
\label{SIArange}
1/3 fm \lesssim |x| \lesssim 1 fm
\ee
Results of simulations including the contribution from all 
instantons and form the leading-instanton only are reported in figure
(\ref{RILvsLI}).
One can see that the agreement is lost for rather large values of $|x|$
($|x| \gtrsim 0.6 fm$).

In the last section we saw that the SIA does  not only assume
  leading-instanton 
dominance, but involves  some effective mass parameters, which 
collectively describe the effects of all other instantons 
(see eqs. (\ref{mstar1}) and (\ref{mstar2})).
In the next section we shall determine numerically 
such parameters.


\section{Numerical study of the Quark Effective Mass Parameters}
\label{mass}
In section \ref{theory} we argued that the universality of the
 effective mass, which collectively describes the effects of all 
non-leading-instantons, can be put in relation to the fluctuations 
of the quark condensates through eq. (\ref{universality}).

Obviously, the accuracy of calculations in the SIA 
depends on the value of $R_m$ (defined in (\ref{Rm})) 
in realistic ensembles.

We have have evaluated $R_m$ and the corresponding effective masses,
numerically
\footnote{Here we have averaged on 5000 configuration of 
256 instantons in a $4^4 fm^4$ box.}
 in the random instanton liquid   and
in the interacting liquid 
(for a review of these   ensembles see \cite{shuryakrev}).
Our results are summarized in table \ref{massresult}:
  \begin{table}
  \caption{Universality parameter, $R_m$ and the effective masses evaluated in
the RILM and in the IILM}
 \label{massresult}
\begin{tabular}{|c|c|c|}\hline
\multicolumn{1}{|c|}{Quantity} & 
\multicolumn{1}{|c|}{RILM calculation} &
\multicolumn{1}{|c|}{IILM calculation} \\ \hline \hline
$R_m$ &   $0.4$  & $0.2$ \\ \cline{2-3}
$m_{uu}$ &  $\ 120 MeV$ & $ 177 MeV$ \\ \cline{2-3}
$\sqrt{m_{uudd}^2}$&$  65 MeV$ & $ 91 MeV$ \\ \hline 
 \end{tabular}
 \end{table}

These results show that, in the instanton vacuum with realistic density,
the {\em universality does not hold} 
\be
\label{inequality}
m_{uu}^2\ne m^2_{uudd}.
\ee
This implies that an effective mass extracted from the quark condesate can
not be used in calculations involving more than one zero-mode propagator.

On the other hand, the results of numerical simulations presented in
section \ref{reliability} have  shown  that matrix elements involving 
only one zero-mode propagator (like the quark condensate) 
can not be reliably evaluated in the SIA, simply  because the leading 
instanton is not dominant.
As a consequence, one is forced to consider only
correlation functions involving at least two 
such propagators and therefore  $m_{uu}$ is of no practical usefulness.

In more general terms, one may address the question
whether the effective mass parameter
depends on the particular correlation 
function being evaluated.
If so, this feature would spoil much of the predictive power 
of the SIA.
In such pessimistic scenario
the  SIA  would only allow 
to work out the functional expressions of small-sized correlations, 
but not their overall normalization.
However we will show that 
the effective mass parameters depend essentially on 
the number of zero-mode propagators involved, and that 
 $m^2_{uudd}$ is in a way universal for a number of applications.
In this case, SIA is predictive including the normalization.
To check that 
we have extracted 
$m^2_2$ 
from the analysis of 
several hadronic two-point functions evaluated in SIA and in the liquid.
In particular, we considered the pion pseudo-scalar the scalar diquark and the 
a nucleon scalar correlation functions:
\be
P(x) &=& <0| J_5(x) J^\dagger_5 (0)|0>,\\
D(x) &=& <0| J^a_{C5}(x) J^{a\,\dagger}_{C5} (0)|0>,\\
N(x) &=&<0| Tr\,[ \eta (x)\,\bar{\eta}(0) \gamma_4 ]|0>, 
\label{proton}
\ee          
where,
\be
J_5(x) &:=& \bar{u}(x)\,\gamma_5 \,d(x).\\
J^a_{C5}(x) &:=& \epsilon^{a\,b\,c}u_b(x)\,C \gamma_{5} \,d_c(x).\\
\eta_\alpha (x) &:=& \epsilon^{a\,b\,c} 
( u^a(x) C\gamma_5 u^b(x)) u_\alpha^c(x).
\ee
All these correlations function are known to receive contribution
from two propagators in the zero-mode.

\begin{figure}[h|t]
 \includegraphics[height=7.0cm,width=8.0cm]{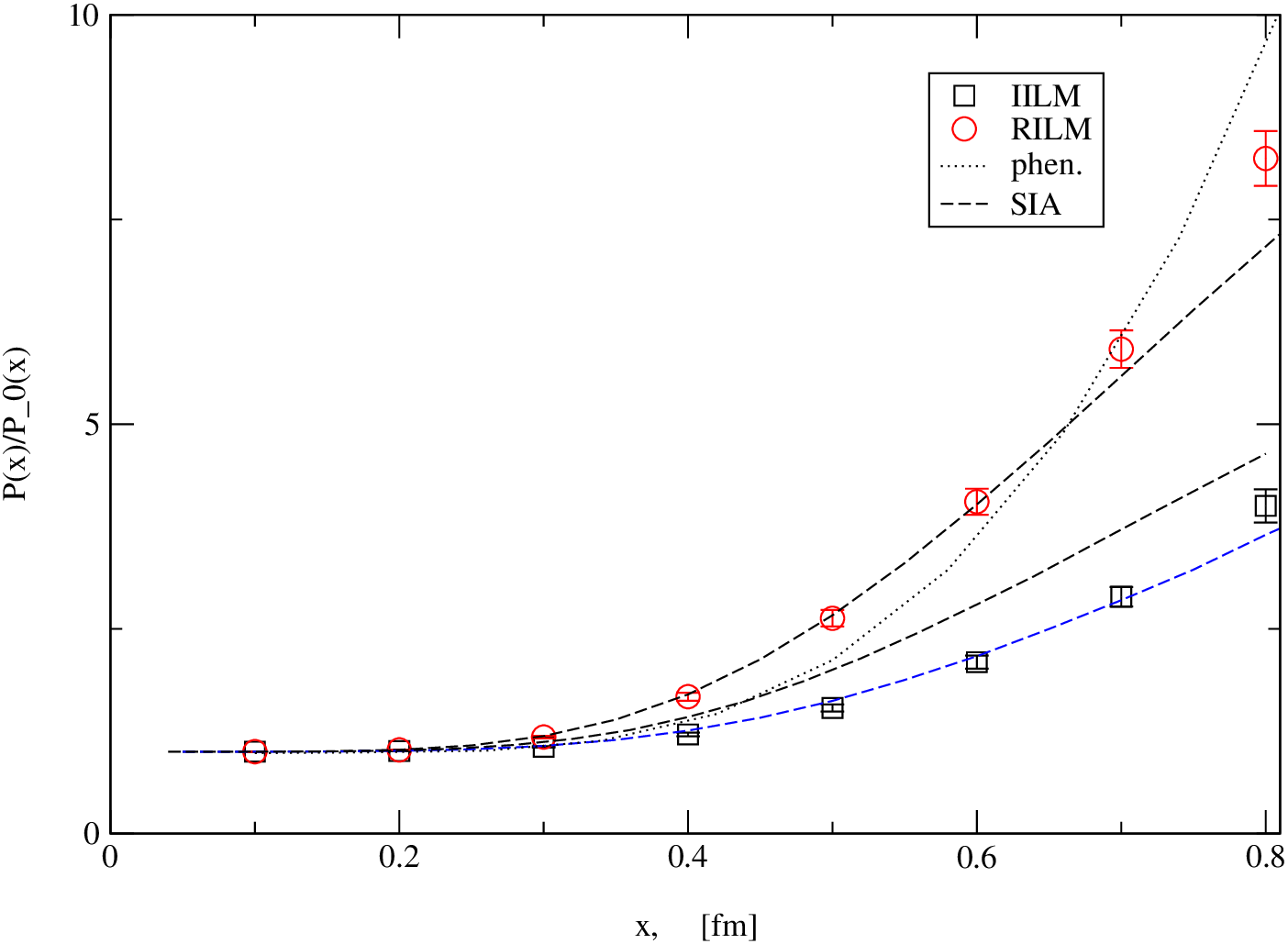}
 \caption{Pion pseudo-scalar two-point function 
normalized to the same correlation function in the free theory.
The open circles (squares) represent RILM (IILM) points, the
 dashed lines represent SIA calculations with masses given in table 
\ref{manymasses} and the dotted line
 is the phenomenological curve obtained from the spectral decomposition.}
 \label{figps2point}
 \end{figure}

\begin{figure}[h|t]
 \includegraphics[height=7.0cm,width=8.0cm]{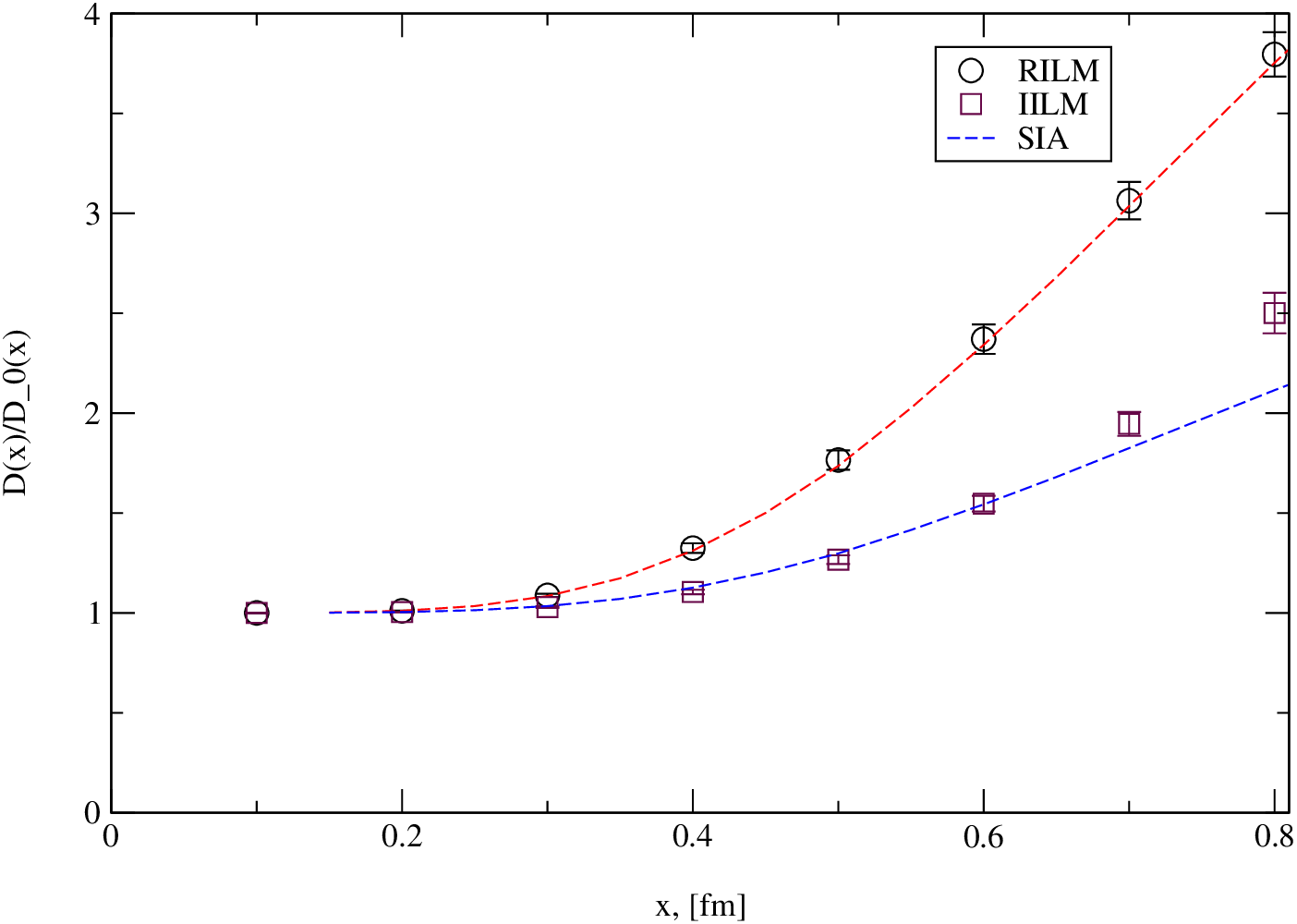}
 \caption{Diquark scalar two-point function 
normalized to the same correlation function in the free theory.
The open circles (squares) represent RILM (IILM) points and the
 dashed lines represent SIA calculations with the effective 
masses given in table \ref{manymasses}}.
 \label{figdiquark2point}
 \end{figure}

\begin{figure}[h|t]
 \includegraphics[height=7.0cm,width=8.0cm]{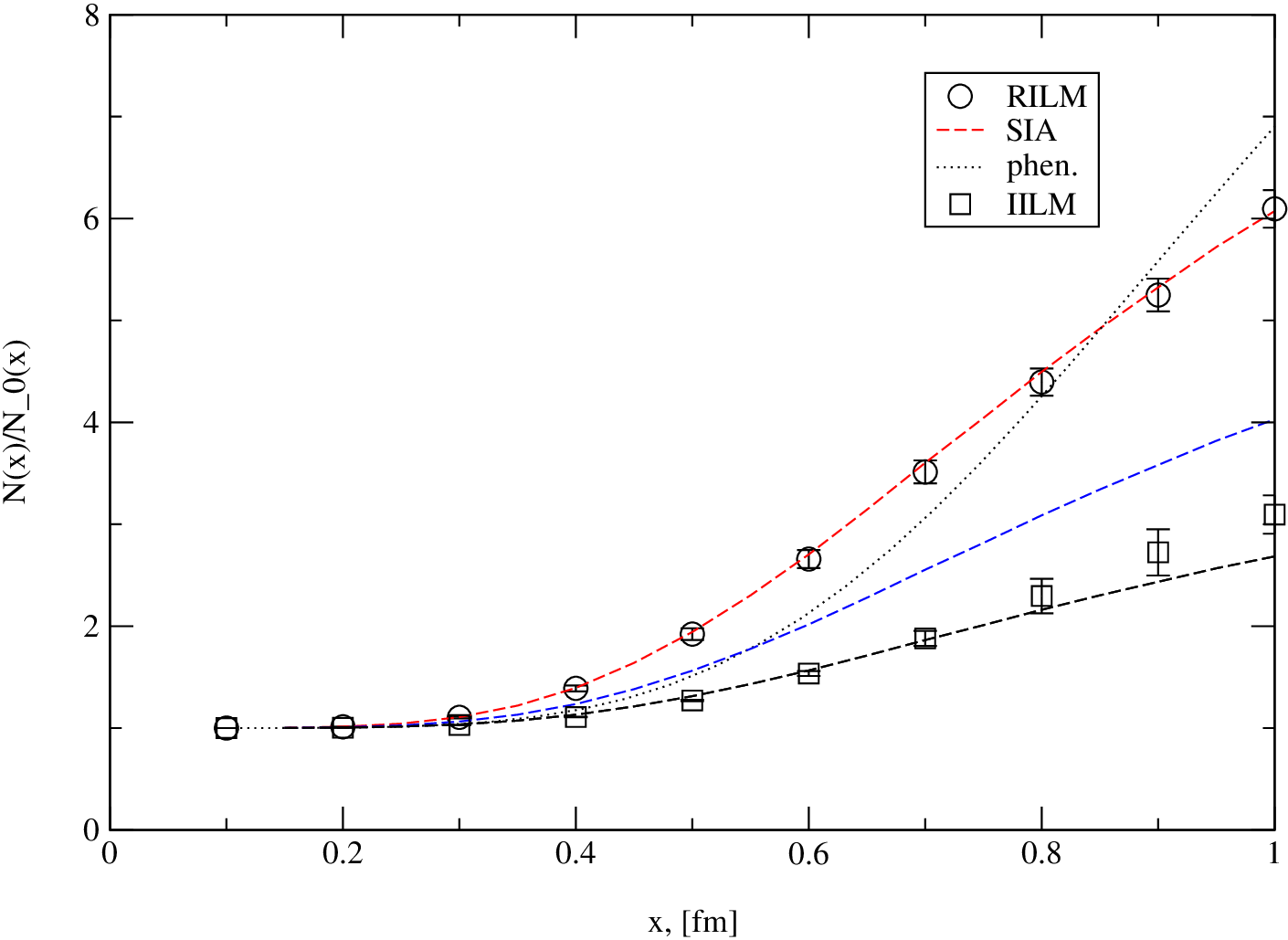}
 \caption{Nucleon scalar two-point function normalized to the same 
correlation function in the free theory.
The open circles (squares) represent RILM (IILM) points and the
 dashed lines represent SIA calculations with the effective 
masses given in table \ref{manymasses}.}
 \label{fignucleon2point}
 \end{figure}

The comparison between results obtained in the SIA, in the random instanton
liquid model (RILM) and in the interacting instanton liquid model (IILM)
 are reported in  figs. (\ref{figps2point}), 
(\ref{figdiquark2point}) and 
(\ref{fignucleon2point}).
The corresponding values for $\sqrt{m_2^2}$ are presented in
table (\ref{manymasses}).
These
values are indeed rather different from
 the traditionally adopted estimate 
$m^*=170 MeV$, extracted from the quark condensate.

  The general reason why these masses are rather small is the
following.
Instantons have fluctuating strength of interaction with others in the 
ensemble: some of them are ``hermits'' and have small matrix
elements
in the corresponding entries of the
overlap matrix $T$. As in all expressions we average the {\em inverse} 
of this
matrix, the contribution of such ``hermits'' is enhanced.
 This lowers the value of the effective masses.
Furthermore, because random ensemble of RILM has more 
such ``hermits'', as compared to
IILM (where the fermionic determinant in the statistical weight 
suppresses them), these masses are smaller in RILM as compared to IILM. 
Such discrepancy reflects the fact that the two ensembles give 
actually quite different correlation functions \cite{shuryakrev}.

  \begin{table}
  \caption{Estimates of the quark effective mass $m^2_2$ from several correlation functions.}
 \label{manymasses}
\begin{tabular}{|c|c|c|}\hline
\multicolumn{1}{|c|}{Correlation function} & 
\multicolumn{1}{|c|}{ $\,m_2^2 \,[MeV^2]\,$  (RILM)} &
\multicolumn{1}{|c|}{ $\,m_2^2 \,[MeV^2]\,$  (IILM)} \\ \hline \hline
$\chi_{uudd}$ condensate &   $ ( 65 )^2$  & $( 91 )^2$\\ \cline{2-3}
pion pseudo-scalar 	 &   $ ( 65 )^2$  & $( 105 )^2$\\ \cline{2-3}
diquark scalar     	 &   $ ( 69 )^2$  & $( 105 )^2$\\ \cline{2-3}
nucleon scalar     	 &   $ ( 67 )^2$  & $( 105)^2$\\ \hline 
 \end{tabular}
 \end{table}

From these results we conclude that $m_2^2$  seems to be a 
universal parameter, describing the collective many-instanton effects.

It is important to know what value of $m_2^{\,2}$ is suggested
by the available phenomenology.
As before, we chose to consider the pion 
pseudo-scalar correlator, because it receives maximal contribution from 
instanton zero-modes.
The traditional ``pole-plus-continuum'' model for the spectral 
decomposition of $P(x)$, gives \cite{meson2pt}, \cite{shuryakcorr}:
\be
\label{pssumrule}
P(x)= \lambda_\pi^2 D(m_\pi;x) + \frac{3}{8\,\pi^2}\int_{s_0}^\infty
\, ds\, s D(\sqrt{s};x),
\ee
where  $D(m;x)$ is the scalar propagator,
$s_0$ is the threshold for the continuum ( $\sqrt{s_0} \simeq 1.6 GeV  $) and
the pseudo-scalar decay constant $\lambda_\pi$ is given by:
\be
\lambda_\pi =<0| \bar{u}\gamma_5d|\pi>= \frac{f_\pi\,m_\pi^2}{m_u+m_d}
\simeq (480 MeV)^2.
\ee

 \begin{table}
  \caption{Determinantal masses evaluated in the RILM and in the IILM as 
compared to $m_1$ and $m_2^2$, defined in section \ref{mass}.}
 \label{determinantalmass}
\begin{tabular}{|c|c|c|}\hline
\multicolumn{1}{|c|}{mass} & 
\multicolumn{1}{|c|}{RILM calculation} &
\multicolumn{1}{|c|}{IILM calculation} \\ \hline \hline
$m_1$ & $120 MeV $ & $177 MeV$ \\ \cline{2-3}
$m_{det}$ & $ 63 MeV   $ & $102 MeV$ \\ \hline
$ m_2^2 $ & $ (65 MeV)^2  $ & $(103 MeV)^2$ \\ \cline{2-3}
$m_{det}^2$ 
&$  (64 MeV)^2 $& $(103 MeV)^2$ \\ \hline
 \end{tabular}
 \end{table}

We determined $m_2^2$, by fitting the  SIA prediction 
to the phenomenological curve  
obtained from (\ref{pssumrule}).
We found (see fig. \ref{figps2point}): 
\be
\label{m2phen}
m_{2\, phen.}^2= (86 MeV)^2.
\ee
To further check the approach, 
we have evaluated the scalar proton two-point function $N(x)$, using the
 value (\ref{m2phen}) and we have compared with the phenomenological curve 
(see fig. (\ref{fignucleon2point})
\footnote{Notice that, in this case, the pole contribution depends on the 
nucleon  ``decay-constant'' $\Lambda_s$, which is 
not known experimentally. We have therefore used an 
estimate ($\Lambda_s=2.5 fm^{-3}$) which reasonably agrees 
with several model calculations
\cite{shuryakrev} .}).
In summary: with this value
 we obtained good very good agreement with phenomenology 
and therefore we suggest that (\ref{m2phen}) should be used f
the  applications of the SIA, when two zero-mode propagators are involved.

\section{Evaluation of an effective mass in the fermionic determinant}
\label{determass}

The propagator is not the only place where the Dirac operator appears:
the QCD statistical sum contains its $determinant$, appearing in power
given by the number of light quark flavors $N_f$. If one considers
the academic vacuum with only one instanton, this determinat
contains the product of ``current'' quark masses for all
quarks \cite{thooftzm1}.
If this would be the final answer for the instanton density,
instanton effect would be strongly suppressed by their small values. 

However, in physical
vacuum there are sufficiently many instantons to break chiral
symmetry and produce non-zero quark condesates and effective quark
masses,
which substitute for much smaller ``current'' masses and make instanton
effects significantly stronger. The interplay between these effective
masses 
and current quark masses is especially interesting for strange quark,
since the former and the latter, $m_s$, are of comparable magnitude.
This issue has been studied e.g. in recent paper \cite{Musakhanov},
where it was concluded that the usual additive formula for total
effective quark mass of the strange quark
$M_s^{tot}=m_{eff}(m_s=0)+m_s$ is wrong, and the true value
 of
$M_s^{tot}$ is not very different from that for u,d quarks because
$m_{eff}(m_s)$ strongly decreases with $m_s$. 

Apart of the role of strange quark mass in general, there is also a
general issue of correct connection of units and  vacuum parameters
(with the instanton density being one of them) for QCD with different
number of flavors. (For example, between no-quark or quenched QCD and
the physical world.) 
In order to study all of this, it important to know what is the
absolute
magnitude of the fermionic determinants in the
instanton-based vacuum models considered. Some of those are repoted in 
this section.

In the instanton-based model context, the fermionic determinant
is usually represented by the determinant of the overlap matrix $T$
(see description e.g. in \cite{shuryakrev}) in the zero mode subspace. 
After averaging over appropriate ensemble, one can
 define the so-called ``determinantal masses'':
\be
m_{det}^i:=
\frac{<(\,det [\ds{D}]\,)^{i/N}\rho^i >}{<\rho^i>},\qquad i=1,2,...
\ee
where index $i$ refers to number of flavors and
$N$ denotes the number of instantons.
Their values tell us how much the presence of fermions reduces
the instanton density, compared to the same ensemble without them.

Originally, in \cite{shifman80}, \cite{shuryak86} 
an estimate for the determinant
  effective mass was estracted from the averaging of the 't Hooft Lagrangian 
assuming factorization of quark condesates, and using the same
$m^*=170 \, MeV$. If so, each flavor reduces instanton density
by the factor $m^*\bar \rho\approx 0.28$.
As we will see shortly, the corresponding reduction
factor is actually even smaller.

In principle, there is no reason why
the values of  $m_{det}$ and  $m_{det}^2$ should agree 
  with $m_1$ and $m_2^2$, defined in the previous section:
we now averahe positive rather than negative powers of the
overlap matrix.

We have evaluated the determinantal masses 
in the RILM and in the IILM.
Results are reported in table \ref{determinantalmass}.
Some comments are in order.
First of all note that, in both ensembles, the values of 
$m_{det}^2$ turn out to be quite consistent
with the values of $m_2^2$.
Furthermore, the fluctuations of the determinantal mass,
$m_{det}$ and $(m_{det})^2$ are very small:
\be
m_{det}^2- (m_{det})^2 \ll m_2^2 - (m_1)^2,
\ee
implying essentially that $m_1$ is inconsistent with $m_{det}$.
This fact could have two possible explanations.
On the one hand, one could argue that $m_1$ is a somewhat 
ill-defined parameter, because the SIA can not be used to
 evaluate quark condensate.
On the other hand, one could observe that larger fluctuations
for the effective masses defined
in \ref{mass} should not be surprising, since such parameters appear
always in denominators  of SIA calculations.


\section{Conclusions and Outlook}
\label{conclusions}

Summarizing our study of the SIA approximation
in QCD, we first notice that this
 approach has been related to the theory of the full 
ensemble and all the effective parameters
previously loosely called ``effective masses'' are defined.
All of them describe different aspects of
  collective interaction between the
``leading'' instanton (the closest to the observation points) and
 all others, and related to the overlap matrix $T$ .
Different effective mass values simply follow from different 
ensemble averaging. In particular,
 the factor $\frac{1}{m_1}$, appearing in SIA calculations 
with one
propagator in the zero-mode, does not correspond to the square root of the
factor $\frac{1}{m^2_2} $, appearing when two such propagators
 are involved. 

We have made numerical simulations in the RILM and IILM  and found    
that the contribution of the leading-instanton  actually dominates all
condensates of  operators of dimension six or more, as well as
short-distance correlation functions ($|x|\lesssim 0.6 fm$).
This however is true  only  for correlation functions with 
at least two zero-mode propagators involved.
Earlier estimates extracted from the quark
condensate are not accurate.

Furthermore, the parameter
 $\frac{1}{m^2_2}$ is  approximatively universal
 for several correlation functions  with \emph{two} 
zero-mode propagators involved.
We have also extracted a phenomenological estimate 
of its value from the analysis of the
pion pseudo-scalar correlator.
We found $m_{2\, phen.}^2 \simeq (86 MeV)^2$, much
smaller than the  value originally obtained from the quark condensate. 
Our new value should
be used in many applications of the SIA.
 
Finally, we have compared our estimates for the effective mass parameters 
$m_1$ and $m_2^2$, with
the measurements of the ``determinantal'' masses,
 introduced in \cite{shuryak86}.
We observed substantial agreement between $m_2^2$ and $m_{det}^2$ 
  both in the RILM  and the IILM, but different
from $m_1$ extracted from the quark condesate alone.
This implies that light quarks are about twice more effective
(per flavor) in diluting
the instanton vacuum density.

\begin{acknowledgments}
We would like to thank H.Forkel whose questions initiated this work, and
T. Sch\"afer for many helpful discussions and numerical help.
The work is partly supported by the US DOE grant No. DE-FG02-88ER40388.
\end{acknowledgments}


\end{document}